\documentclass[twocolumn]{aastex63}
\usepackage{graphicx}
\usepackage{hyperref}
\usepackage{epstopdf}

\received{December 19, 2019}
\revised{January 20, 2020}
\accepted{January 22, 2020}
\begin{document}

\title{A Word to the \emph{WISE}: Confusion is Unavoidable for \emph{WISE}-selected Infrared Excesses} 

\correspondingauthor{Erik Dennihy}
\email{edennihy@gemini.edu}

\author[0000-0003-2852-268X]{Erik Dennihy}
\affil{NSF's National Optical-Infrared Astronomy Research Laboratory \\ Gemini Observatory, Colina el Pino S/N, La Serena, Chile}

\author[0000-0003-1748-602X]{Jay Farihi}
\affil{Department of Physics and Astronomy, University College London, London WC1E 6BT, UK}

\author[0000-0002-6428-4378]{Nicola Pietro Gentile Fusillo}
\affil{Department of Physics, University of Warwick, Coventry, CV4 7AL, UK}
\affil{European Southern Observatory, Karl-Schwarzschild-Str 2, D-85748 Garching, Germany}

\author[0000-0002-1783-8817]{John H. Debes}
\affil{ESA for AURA, Space Telescope Science Institute, 3700 San Martin Drive, Baltimore, MD 21218, USA}

\begin{abstract}
Stars with excess infrared radiation from circumstellar dust are invaluable for studies of exoplanetary systems, informing our understanding on processes of planet formation and destruction alike. All-sky photometric surveys have made the identification of dusty infrared excess candidates trivial, however, samples that rely on data from \emph{WISE} are plagued with source confusion, leading to high false positive rates. Techniques to limit its contribution to \emph{WISE}-selected samples have been developed, and their effectiveness is even more important as we near the end-of-life of \emph{Spitzer}, the only facility capable of confirming the excess. Here, we present a \emph{Spitzer} follow-up of a sample of 22 \emph{WISE}-selected infrared excess candidates near the faint-end of the \emph{WISE} detetection limits. Eight of the 22 excesses are deemed the result of source confusion, with the remaining candidates all confirmed by the \emph{Spitzer} data. We consider the efficacy of ground-based near-infrared imaging and astrometric filtering of samples to limit confusion among the sample. We find that both techniques are worthwhile for vetting candidates, but fail to identify all of the confused excesses, indicating that they cannot be used to confirm \emph{WISE}-selected infrared excess candidates, but only to rule them out. This result confirms the expectation that \emph{WISE}-selected infrared excess samples will always suffer from appreciable levels of contamination, and that care should be taken in their interpretation regardless of the filters applied. 
\vspace{0.5cm}
\end{abstract}

\section{Introduction}

In the era of database astronomy, the construction of spectral energy distributions (SEDs) from the ultra-violet to the mid-infrared for large samples of stars is straightforward, requiring little user-input or effort. Modern tools such as the VO SED Analyzer \citep{bay08:aa492} can even detect infrared excesses for thousands of candidates at a time in a completely automated fashion. Many sub-fields have benefited from the ease-of-use of catalog photometry, though they are not without pitfalls. Searches for infrared excesses from warm (1000\,K), circumstellar dust provide a good case-study of the benefits and drawbacks of analyzing SEDs using only catalog photometry. 

Circumstellar dust is a signpost for planetary systems, indicating the on-going process of planetary formation around pre-main and main-sequence stars \citep{ken12:mnras426,pat14:apjs212,cot16:apjs225,bin17:mnras469}, and illuminating the post-main sequence destruction of remnant planetary systems around white dwarf stars \citep{deb11:apjs197,hoa13:apj770,bar14:apj786,den17:apj849}. The frequency of circumstellar dust around stellar sources informs planetary occurrence rates in instances where direct detection is not feasible. These searches rely heavily on data from the \emph{Wise Infrared Survey Explorer} (\emph{WISE}; \citealt{wri10:aj140}), which produced the only all-sky survey at the wavelengths where warm dust is most apparent ($\lambda\geq$\,3\,$\mu$m). But the coarse spatial resolution of \emph{WISE} leads to a high probability of source confusion, contaminating samples of \emph{WISE}-selected infrared excess with false positives and skewing statistical studies of warm dust frequency. 

Estimates of contamination by source confusion for \emph{WISE}-selected dusty infrared excesses around main-sequence stars indicate false-positive rates as high as 70\% \citep{sil18:apj868}. Dusty infrared excesses around white dwarf stars are much fainter than their main-sequence counterparts, and typically only detected in \emph{W1} and \emph{W2} bands (see \citealt{far16:nar71} for a recent review). Their faint magnitudes push the boundaries of the source-confusion limited detection thresholds of the AllWISE surveys. More concerning, as the \emph{Spitzer Space Telescope} \citep{wer04:apjs154} reaches its end-of-life, the ability to confirm \emph{WISE} infrared excesses for large samples may be lost entirely. The effectiveness of the next generation observatory, the \emph{James Webb Space Telescope} \citep{gar06:ssr123}, to mimic the survey imaging capability of \emph{Spitzer} will be limited by initial slew times that are an order of magnitude larger. This is likely to mean \emph{JWST} cannot support this science effectively for large samples of dusty white dwarfs as are currently being identified with \emph{Gaia} \citep{reb19:mnras489}.

\begin{figure*}[t!]
\gridline{\fig{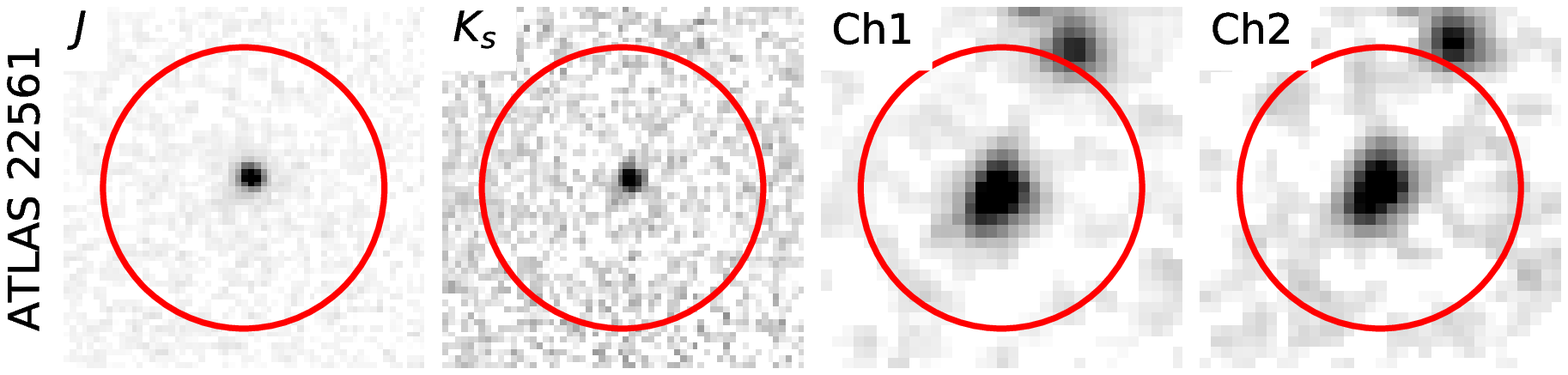}{0.495\textwidth}{ }\fig{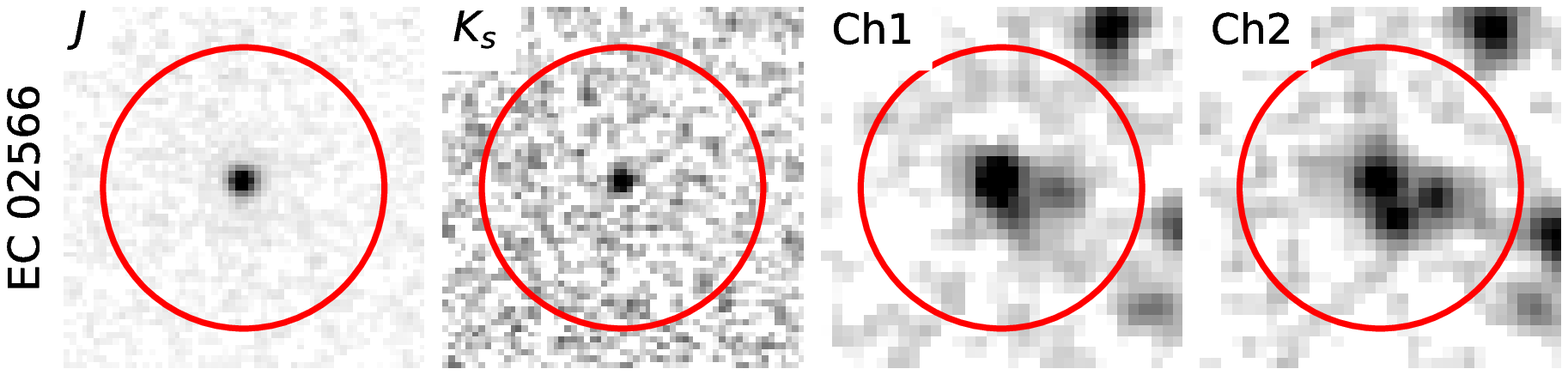}{0.495\textwidth}{}}
\vspace{-1.2cm}
\gridline{\fig{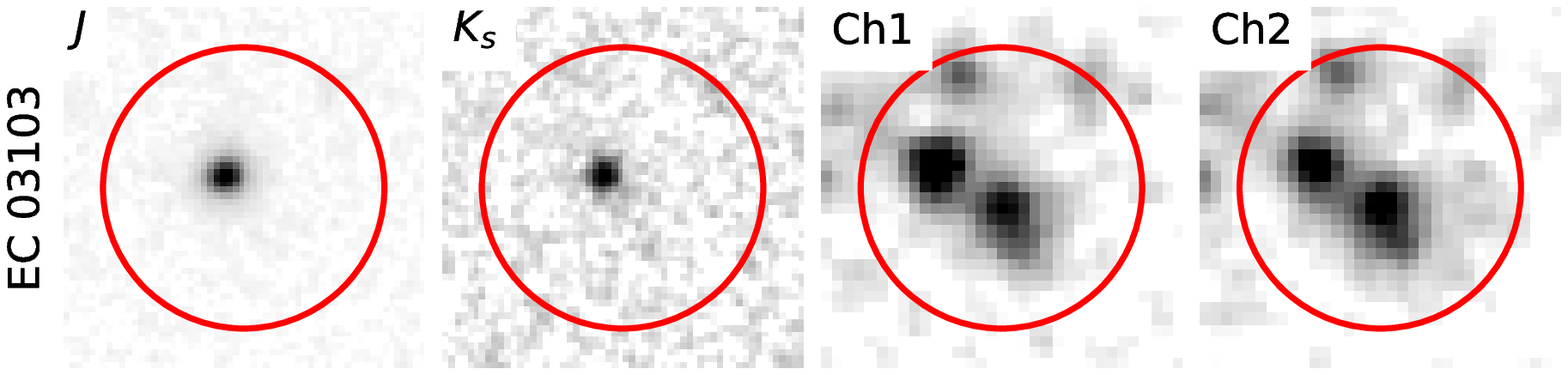}{0.495\textwidth}{}\fig{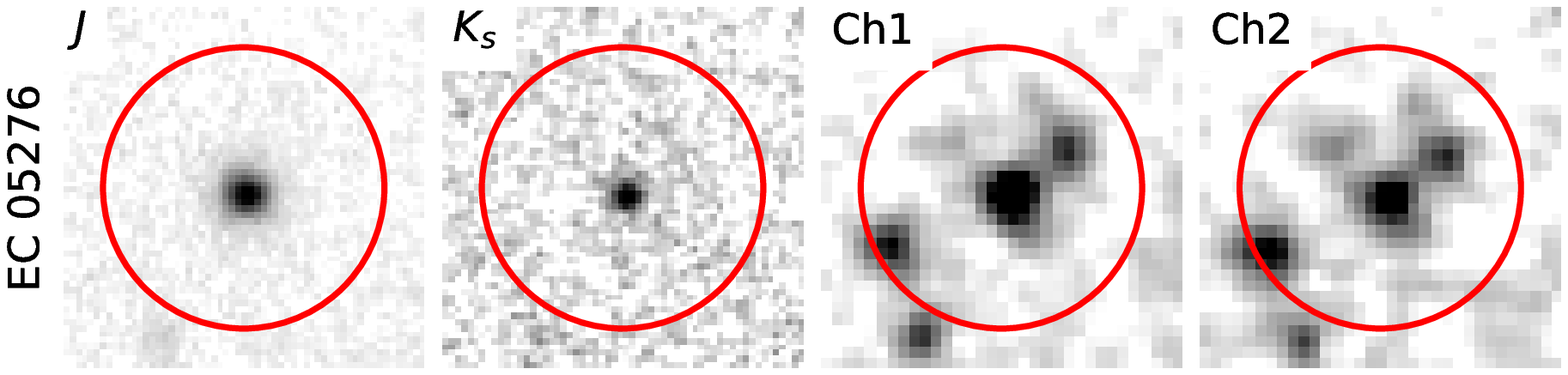}{0.495\textwidth}{}}
\vspace{-1.2cm}
\gridline{\fig{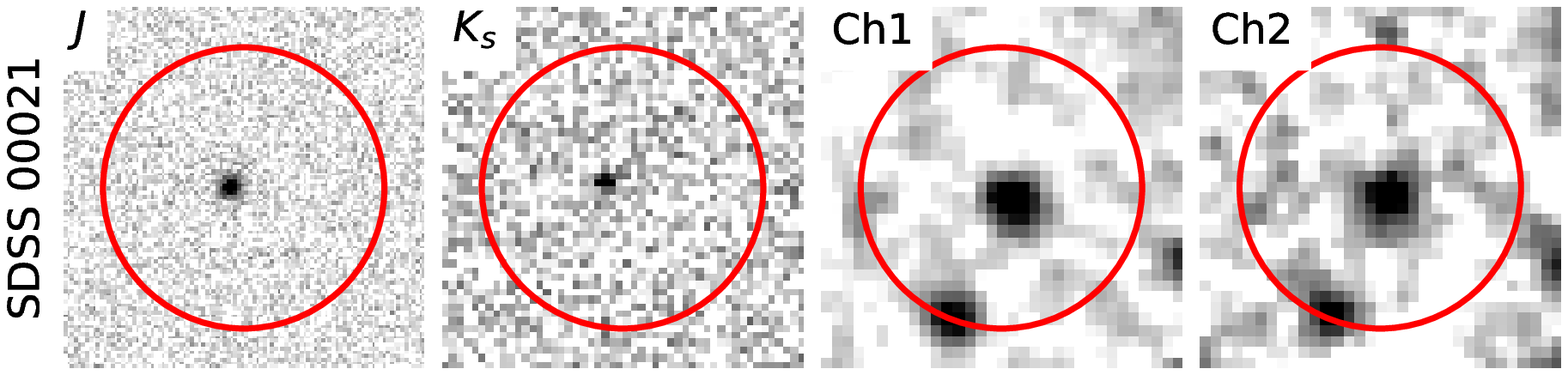}{0.495\textwidth}{}\fig{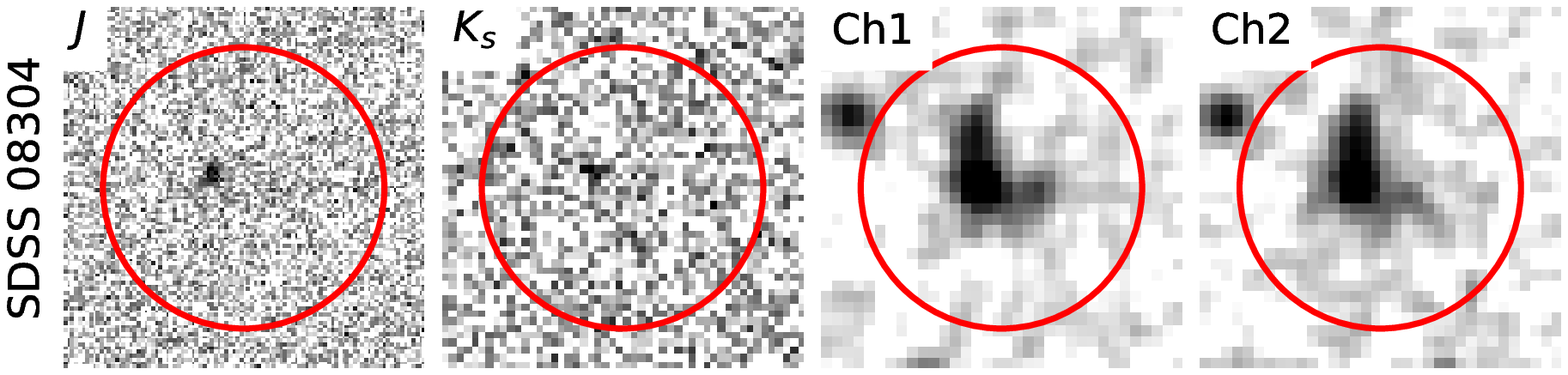}{0.495\textwidth}{}}
\vspace{-1.2cm}
\gridline{\fig{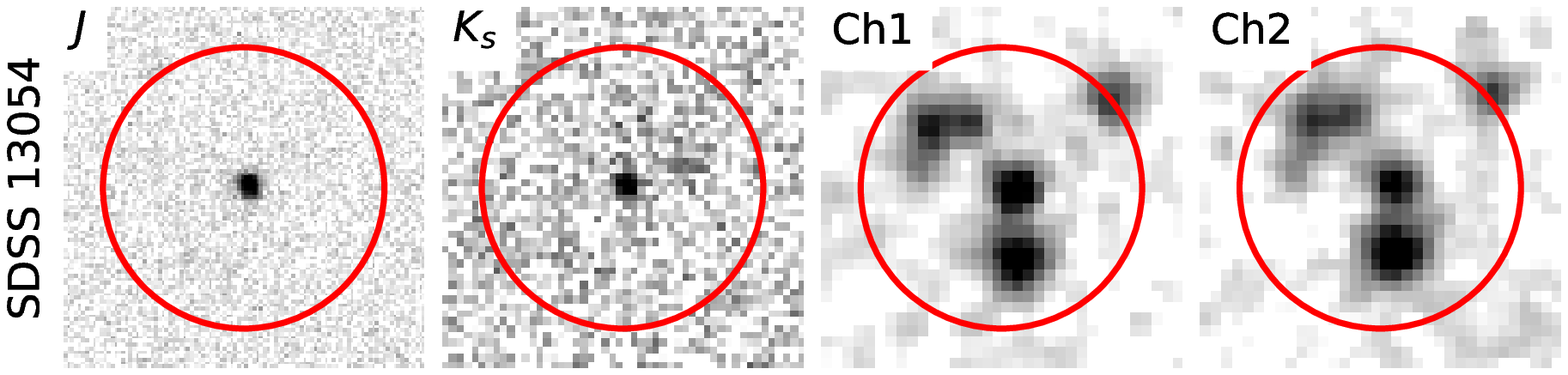}{0.495\textwidth}{}\fig{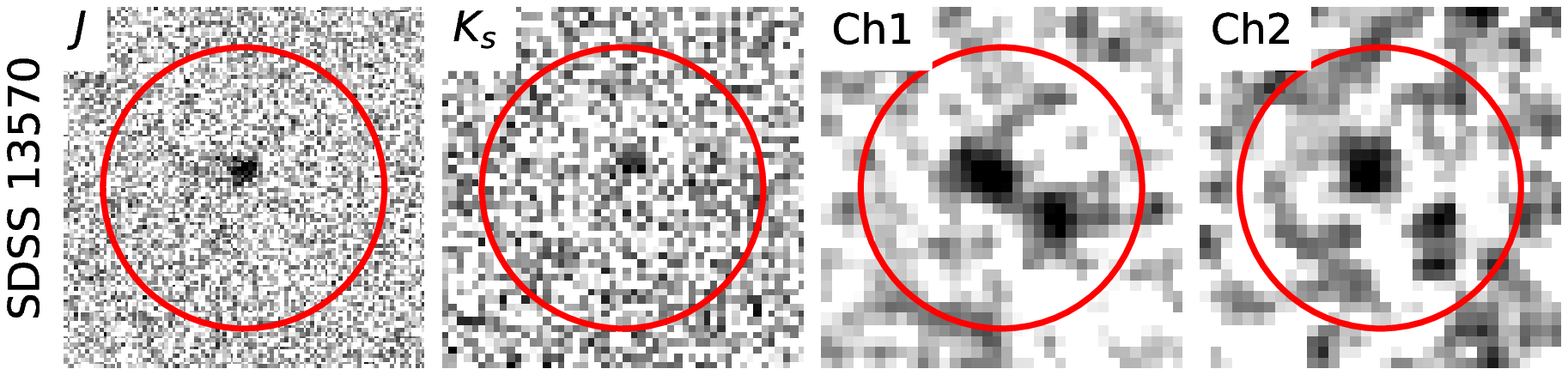}{0.495\textwidth}{}}
\vspace{-0.5cm}
\caption{\added{Ground-based near-infrared and \emph{Spitzer} imaging of the eight false positive candidates is shown in order of increasing wavelength}. From left to right we show \emph{J}, \emph{K}$_s$, IRAC-Ch$\,$1 and IRAC-Ch$\,$2 images centered on the AllWISE source position for each target with a 7.8\arcsec\, circle over-plotted to visualize the \emph{WISE} beam size (1.3\,$\times$\,FWHM in \emph{W1}) \added{as a proxy for the source confusion limit}. The AllWISE pipeline includes an active deblending routine that can resolve up to two sources within this separation, but none of our targets (including those not shown here) were flagged for active deblending.\added{The images succinctly demonstrate that near-infrared imaging is insufficient to rule out source confusion in the \emph{WISE} \emph{W1} and \emph{W2} bands}\label{fig:imsequence}} 
\end{figure*}

In this paper, we present \emph{Spitzer} follow-up of a sample of 22 \emph{WISE}-selected infrared excess candidates around white dwarf stars and discuss the efficacy of techniques to limit the contamination of \emph{WISE}-selected infrared excesses by source confusion. This sample approaches the faint limit of the AllWISE surveys, making it of broader impact to studies of source confusion amongst \emph{WISE}-selected infrared excesses. Using the higher-resolution \emph{Spitzer} data, we confirm the \emph{WISE} infrared excesses in 14/22 systems, with the remaining systems all showing nearby sources within the \emph{WISE} beam. 

Prior to their \emph{Spitzer} observations, all of our targets were vetted by examining ground-based near-infrared imaging and astrometric shifts to probe for clear instances of confused \emph{WISE} photometry. None of the eight contaminated systems showed nearby sources in their ground-based near-infrared imaging, demonstrating that it is insufficient to rule out source confusion at the \emph{WISE} bands. We find that the astrometric information is a more useful indicator of the potential for source confusion, but only when considering the full astrometric uncertainty of the surveys involved. Even when applied carefully, we demonstrate that these techniques will not result in clean sample of excesses, and studies based on \emph{WISE}-selected infrared excesses should always consider a level contamination when interpreting sample properties. 

\section{\emph{Spitzer} View of \emph{WISE} Infrared Excess Candidates}

Our targets were selected from a handful of studies that applied different criteria to identify the infrared excesses (\citealt{den17:apj849}, Gentile-Fusillo et al. in prep). The common property of our targets is an infrared excess in the \emph{WISE} \emph{W1} and \emph{W2} consistent with a warm, compact dust disk around a white dwarf star. The \emph{Spitzer} photometry is superior to the \emph{WISE} photometry in both sensitivity and, more importantly, spatial resolution, allowing us to test the possibility that a given \emph{WISE} excess is the result of source confusion. For each target, we searched for instances of multiple sources within the \emph{WISE} beam, and compared the \emph{Spitzer} photometry against stellar models to confirm the \emph{WISE}-selected excess.

\subsection{IRAC Imaging and Photometry}

Under program 14100, we collected 3.6 and 4.5 $\mu$m photometry of 22 dusty white dwarf candidates using the Infrared Array Camera (IRAC; \citealt{faz04:apjs154}) with \emph{Spitzer} in Cycle 14. Ten frames were taken using 30\,s exposures with the medium-sized cycling dither pattern, resulting in 300\,s of total integration in each channel. We produced fully calibrated mosaic images for each target using the MOPEX software package \citep{mak06:spie6274} following the recipes outlined for point-source extraction in the \emph{Spitzer} Data Analysis Cookbook version 6.0. PSF-fitted photometry was conducted using APEX, and the error in the measured flux was summed in quadrature with a 5\% calibration uncertainty \citep{far08:apj674}. It has been demonstrated that well-dithered observations are robust against intra-pixel flux variations at the sub-percent level \citep{wil19:mnras487} so we did not apply any such corrections. The measured fluxes are presented in Table \ref{tab:excesstable}. 

For each target, we examined the IRAC-Ch$\,$1 and Ch$\,$2 mosaic images for multiple sources within the \emph{WISE} beam, centered on the AllWISE detection. The critical distance for resolving neighboring sources is 1.3$\,\times\,$the full-width half-maximum of the point-spread function of a given band (7.8\arcsec\,for \emph{W1}). Within this separation, the AllWISE pipeline relies on an active deblending procedure to detect instances of source confusion, triggered by an unsatisfactory fit to the intensity distribution during the point-source fitting photometry routine\footnote{\url{http://wise2.ipac.caltech.edu/docs/release/allsky/expsup/sec4_4c.html}}. None of our targets were flagged for the active deblending routine so we adopted a 7.8\arcsec\,radius as our limit for potential source confusion.

Eight of our 22 targets have multiple sources within this limit indicating the AllWISE photometry was potentially confused. In Figure \ref{fig:imsequence}, we show 10\arcsec$\times$10\arcsec\, cutouts of the publicly available near-infrared \emph{J} and \emph{K}$_{s}$ and IRAC-Ch$\,$1 and Ch$\,$2 images of these eight targets. We note that in all eight, the nearby contaminants are not detected in any of the near-infrared images. We discuss the  efficacy of near-infrared imaging for limiting contamination in \emph{WISE}-selected samples in Section \ref{subsec:nir}.

\subsection{Comparison with Stellar Models}

We constructed SEDs for each target utilizing data from the \emph{Galaxy Evolution Explorer} (\emph{GALEX}; \citealt{mar05:apj619}), Sloan Digital Sky Survey \citep{ahn14:apjs211}, VST-ATLAS survey \citep{sha15:mnras451}, Panoramic Survey Telescope and Rapid Response System \citep{cha16:arxiv}, SkyMapper Southern Survey \citep{wol18:pasa35}, UKIRT Infrared Deep Sky Survey \citep{law07:mnras379}, VISTA Hemisphere Survey (VHS; \citealt{irw04:spie5493, ham08:mnras384, cro12:aap548}), and the AllWISE surveys \citep{cut13:ycat2328}. We de-reddened the photometry using a standard prescription \citep{gen19:mnras482} and converted the magnitudes into fluxes using the published zero-points for each bandpass. 

Most of the objects in our sample do not have a published spectrum to help us choose an appropriate stellar model, instead have only been classified as white dwarf stars. The `EC' objects were first identified with low-resolution spectrograms as part of the Edinburgh-Cape Blue Object Survey \citep{sto97:mnras287}, and later confirmed with targeted follow-up \citep{den17:apj849}. The `ATLAS' and `SDSS' objects were identified as high probability white dwarf candidates via their photometry and proper motion \citep{gir11:mnras417, gen15:mnras448, gen17:mnras469}. All of our objects were also included in the \emph{Gaia} white dwarf catalog of \cite{gen19:mnras482}, which includes estimates of effective temperature and surface gravity assuming both hydrogen and helium dominated atmospheres. 

For our stellar models, we utilized the pure hydrogen-dominated white dwarf model spectra of \cite{koe10:memsai81}, with the effective temperature and surface gravity of each star taken from the hydrogen model fits to the \emph{Gaia} photometry \citep{gen19:mnras482}. It should be emphasized that in our comparion of the model to the SED, the model parameters were not being re-fit to the photometry, rather the surface gravity and effective temperature were fixed and the model was then scaled to fit the optical photometry. Because the goal of this exercise was only to identify the systems with an infrared excess, rather than to fit or describe the infrared excess, this approach was sufficient.

\begin{figure}
\epsscale{1.2}
\plotone{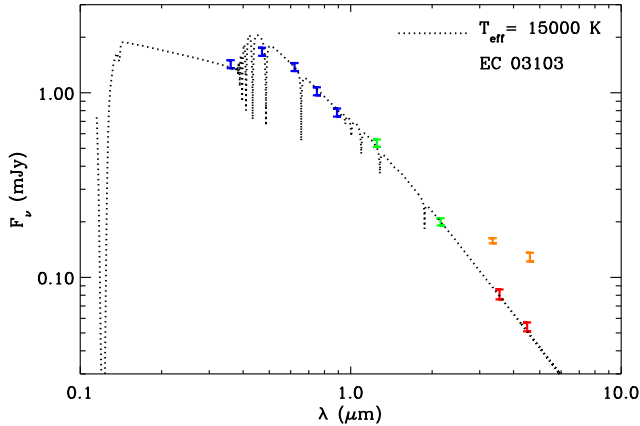}
\caption{The SED of EC\,03103 demonstrates a case of confused \emph{WISE} photometry (orange) erroneously being classified as an excess. The \emph{Spitzer} photometry of the white dwarf (red) is consistent with the stellar model, and the confused sources that produced the \emph{WISE} excess are clearly resolved with \emph{Spitzer} in Figure \ref{fig:imsequence}. The remainder of the spectral energy distributions are shown in Appendix \ref{appendix}. \label{fig:ec03103_sed}}
\end{figure}

We determined the flux excess of each target in the IRAC-Ch$\,$1 and Ch$\,$2 bands using the standard formula:

\begin{equation}
\chi = \frac{F_{\text{obs}}-F_{\text{mod}}}{\sqrt{\sigma_{\text{obs}}^2 + \sigma_{\text{mod}}^2}}
\end{equation}

and deemed those that have a Ch$\,$1 or Ch$\,$2 flux excess greater than 4$\sigma$ and clean IRAC-Ch$\,$1 and Ch$\,$2 images to be \emph{Spitzer} confirmed excesses. Targets that showed IRAC photometry consistent with the stellar model and had multiple sources within our 7.8\arcsec\, confusion limit were the result of confused \emph{WISE} photometry. We present an example SED in Figure \ref{fig:ec03103_sed}, that shows an instance of a contaminated excess produced by source confusion. The contaminating sources for this target, EC\,03103, are clearly identifiable in Figure \ref{fig:imsequence}. The remainder of the SEDs are shown in Appendix \ref{appendix} and in Table \ref{tab:excesstable}, we identify the remaining cases of confused \emph{WISE} photometry. 

Among our sample of 22 targets, we identify eight \emph{WISE}-selected excesses that are the result of source confusion, for a nominal contamination rate of 36\%. It is worth re-emphasizing that our sample had already been vetted for obvious cases of source confusion prior to observation with \emph{Spitzer}, so this 36\% contamination rate only includes cases of source confusion which were \emph{unable} to rule out with ground-based data. The effectiveness of these vetting techniques is discussed below. 

\begin{deluxetable*}{lcccccccc}
\tablecaption{\emph{Spitzer} and \emph{WISE} fluxes for each candidate, separated into \emph{Spitzer} confirmed excesses and confused \emph{WISE} photometry. In the final two columns, we present the \emph{Gaia} Figure of Merit (FoM) and separation from expected position collected from the official \emph{Gaia}-AllWISE cross-match, discussed in Section \ref{subsec:astro}\label{tab:excesstable}}
\tablehead{\colhead{} & \colhead{} & \multicolumn2c{\emph{Spitzer}} & \multicolumn2c{\emph{WISE}} & \colhead{} & \multicolumn2c{\emph{Gaia}}\\ 
\colhead{Target Name} & \colhead{\emph{Gaia} WD Designation\tablenotemark{a}} & \colhead{Ch$\,$1} & \colhead{Ch$\,$2} & \colhead{\emph{W1}} & \colhead{\emph{W2}} & \colhead{S/N} & \colhead{FoM} &\colhead{Separation}\\
\colhead{} & \colhead{} & \colhead{($\mu$Jy)} & \colhead{($\mu$Jy)} & \colhead{($\mu$Jy)} & \colhead{($\mu$Jy)} & \colhead{(\emph{W1})} & \colhead{} & \colhead{(\arcsec)}}
\startdata
\cutinhead{\emph{Spitzer} Confirmed Excess}
ATLAS\,00254  &  WD\,J002540.01--393454.56  &  39$\,\pm\,$3  &  42$\,\pm\,$3  &  34$\,\pm\,$5  &  33$\,\pm\,$9  &  7.3  &  5.5  &  0.66 \\ 
ATLAS\,02325  &  WD\,J023252.01--095745.86  &  49$\,\pm\,$3  &  40$\,\pm\,$3  &  48$\,\pm\,$5  &  41$\,\pm\,$10  &  10.8  &  7.1  &  0.19 \\ 
ATLAS\,10552  &  WD\,J105524.50--023721.13  &  86$\,\pm\,$5  &  86$\,\pm\,$5  &  104$\,\pm\,$7  &  98$\,\pm\,$13  &  16.4  &  0.2  &  0.83 \\ 
ATLAS\,12123  &  WD\,J121236.94--105355.07  &  49$\,\pm\,$3  &  47$\,\pm\,$3  &  43$\,\pm\,$6  &  46$\,\pm\,$12  &  7.8  &  6.2  &  0.47 \\ 
ATLAS\,15131  &  WD\,J151312.71--152352.87  &  35$\,\pm\,$3  &  38$\,\pm\,$3  &  36$\,\pm\,$6  &  30$\,\pm\,$12  &  6.9  &  5.4  &  0.65 \\ 
ATLAS\,22120  &  WD\,J221202.88--135239.96  &  156$\,\pm\,$8  &  156$\,\pm\,$8  &  132$\,\pm\,$7  &  145$\,\pm\,$13  &  19.3  &  8.7  &  0.07 \\ 
ATLAS\,23403  &  WD\,J234036.64--370844.72  &  169$\,\pm\,$9  &  161$\,\pm\,$9  &  155$\,\pm\,$7  &  158$\,\pm\,$11  &  24.4  &  9.4  &  0.05 \\ 
EC\,01071  &  WD\,J010933.16--190117.56  &  95$\,\pm\,$6  &  79$\,\pm\,$5  &  89$\,\pm\,$6  &  99$\,\pm\,$11  &  15.7  &  5.6  &  0.46 \\ 
EC\,01129  &  WD\,J011501.17--520744.67  &  55$\,\pm\,$4  &  34$\,\pm\,$2  &  71$\,\pm\,$6  &  27$\,\pm\,$9  &  14.2  &  6.6  &  0.29 \\ 
EC\,21548\tablenotemark{b}  &  WD\,J215823.88--585353.81  &  199$\,\pm\,$11  &  151$\,\pm\,$8  &  205$\,\pm\,$8  &  171$\,\pm\,$10  &  29.1  &  --  &  -- \\ 
SDSS\,01190  &  WD\,J011909.99+104454.09  &  89$\,\pm\,$5  &  87$\,\pm\,$5  &  90$\,\pm\,$6  &  92$\,\pm\,$11  &  16.0  &  7.5  &  0.22 \\ 
SDSS\,09355  &  WD\,J093553.30+105722.97  &  33$\,\pm\,$3  &  32$\,\pm\,$2  &  37$\,\pm\,$6  &  40$\,\pm\,$12  &  6.5  &  5.0  &  0.85 \\ 
SDSS\,09514  &  WD\,J095144.01+074957.41  &  76$\,\pm\,$5  &  77$\,\pm\,$4  &  65$\,\pm\,$6  &  77$\,\pm\,$12  &  11.2  &  5.5  &  0.56 \\ 
SDSS\,13125  &  WD\,J131251.36+295535.98  &  45$\,\pm\,$3  &  48$\,\pm\,$3  &  38$\,\pm\,$5  &  43$\,\pm\,$10  &  8.4  &  6.6  &  0.31 \\ 
\cutinhead{Confused \emph{WISE} Photometry}
ATLAS\,22561  &  WD\,J225612.92--131938.83  &  91$\,\pm\,$5  &  60$\,\pm\,$4  &  119$\,\pm\,$7  &  74$\,\pm\,$13  &  17.1  &  0.4  &  0.72 \\ 
EC\,02566  &  WD\,J025859.58--175020.33  &  40$\,\pm\,$3  &  22$\,\pm\,$2  &  48$\,\pm\,$5  &  56$\,\pm\,$8  &  11.9  &  3.4  &  0.77 \\ 
EC\,03103  &  WD\,J031121.31--621515.72  &  81$\,\pm\,$5  &  53$\,\pm\,$3  &  157$\,\pm\,$6  &  128$\,\pm\,$8  &  30.4  &  0.0  &  0.42 \\ 
EC\,05276  &  WD\,J052912.10--430334.49  &  71$\,\pm\,$4  &  41$\,\pm\,$3  &  112$\,\pm\,$5  &  65$\,\pm\,$8  &  23.3  &  3.0  &  0.45 \\ 
SDSS\,00021  &  WD\,J000216.18+073350.30  &  30$\,\pm\,$3  &  19$\,\pm\,$2  &  40$\,\pm\,$6  &  37$\,\pm\,$12  &  7.7  &  5.6  &  0.55 \\ 
SDSS\,08304\tablenotemark{c}  &  WD\,J083047.28+001041.51  &  28$\,\pm\,$3  &  27$\,\pm\,$2  &  26$\,\pm\,$6  &  35$\,\pm\,$11  &  5.0  &  0.3  &  2.03 \\ 
SDSS\,13054  &  WD\,J130542.73+152541.16  &  37$\,\pm\,$3  &  23$\,\pm\,$2  &  80$\,\pm\,$6  &  57$\,\pm\,$12  &  14.7  &  5.0  &  0.52 \\ 
SDSS\,13570  &  WD\,J135701.68+123145.62  &  9$\,\pm\,$2  &  6$\,\pm\,$1  &  21$\,\pm\,$5  &  18$\,\pm\,$9  &  5.2  &  4.5  &  0.92 \\ 
\enddata
\tablenotetext{a}{\cite{gen19:mnras482}}
\tablenotetext{b}{The \emph{Gaia}-AllWISE cross-match returned no results for EC\,21548, despite an AllWISE detection within 0.5\arcsec of the expected position. This case is discussed in Section \ref{subsec:astro}.}
\tablenotetext{c}{The measured IRAC-Ch$\,$1 and Ch$\,$2 fluxes are confused with a background galaxy.}
\end{deluxetable*}

\section{Mitigating Contamination in \emph{WISE}-selected Samples}

As \emph{Spitzer} nears the end of its operational lifetime, it is worth considering what techniques are effective at separating the clean from the confused amongst \emph{WISE}-selected infrared excess samples. Recent works have explored this subject using samples of main-sequence stars \citep{pat17:aj153,sil18:apj868}, but the infrared excesses exhibited by dusty debris around white dwarf stars are much fainter, and typically only detected in the \emph{W1} and \emph{W2} bands. Furthermore white dwarf infrared excess searches are often limited to a few dozen candidates, so statistical methods for isolating outliers (such as demonstrated by \citealt{pat17:aj153}) are untenable. In the following sections, we consider a few commonly employed strategies and discuss their effectiveness based our classifications with \emph{Spitzer}. 

\subsection{Ground-based Near-infrared Imaging \label{subsec:nir}}

In the absence of space-based follow-up, ground-based near-infrared imaging can be used to search for instances of multiple sources within the \emph{WISE} imaging beam. The Two Micron All Sky Survey \citep{skr06:aj131} is insufficient in both depth and resolution for these purposes. The UKIDSS Large Area Survey \citep{law07:mnras379} and the VISTA-VHS \citep{mcm13:msngr154} have depths of \emph{K}$\approx18.2$ mag and \emph{K}$_s\approx19.8$ mag, and their images have proven useful for quantifying levels of source confusion (e.g. \citealt{deb11:apj729}, \citealt{den16:apj831}). In the absence of publicly available imaging, targeted programs can also be used to cull samples of \emph{WISE}-selected infrared excesses \citep{bar12:apj760}. Near-infrared imaging is preferred to optical in order to get as close as possible to the bandpass of \emph{WISE} images. Ultracool dwarfs only become apparent beyond 1\,$\mu$m \citep{bar15:aap577} and dusty background galaxies can rise in flux as a power law at the \emph{WISE} wavelengths, escaping detection at optical and even near-infrared wavelengths. 

Prior to their selection for follow-up with \emph{Spitzer}, all 22 of our targets were vetted for nearby sources within the \emph{WISE} beam using high-quality, ground-based near-infrared images. In Figure \ref{fig:imsequence}, we show the \emph{J} and \emph{K}$_s$-band images for the eight contaminated targets. It is apparent from these image sequences that a clean near-infrared image is insufficient to confirm a \emph{WISE}-selected infrared excess candidate. Near-infrared imaging is however a valuable tool for ruling out \emph{WISE}-selected infrared excess candidates in cases where a clear, nearby source can be identified. It should always be considered for vetting candidates when available. 

\subsection{Astrometric Separation\label{subsec:astro}}

Another method to assess the potential for source confusion of a \emph{WISE}-selected infrared excess is to compare its expected position to the detected AllWISE detection. A sufficiently bright and nearby contaminant can be expected to shift the centroid of the detected source in the \emph{WISE} images, indicating source confusion \citep{wil17:mnras468,wil18:mnras481}. 

Prior to their \emph{Spitzer} observations, our candidates were also vetted for large separations between their expected, proper motion-corrected \emph{Gaia} position and their detected AllWISE position. All but one candidate was found within 1\arcsec\, of its proper motion corrected position. The contamination rate in our sample indicates that at the sub-arcsecond level, the raw separation value between the expected and detected position is a poor indicator of source confusion. This can be seen by comparing the separations in Table \ref{tab:excesstable}, where there is a large scatter and overlap between the confirmed and confused samples. The cause of this scatter is the wide range of \emph{WISE} astrometric uncertainty among our targets, and is a by-product of our sample being near the fainter end of of the AllWISE detection limits. Incorporating this astrometric uncertainty is essential for discriminating clean and confused \emph{WISE} photometry, as discussed below.

\subsection{The Gaia Figure of Merit as a Confusion Discriminant}
The astrometric uncertainty of \emph{WISE} is known to be inversely proportional to the detection's signal-to-noise (S/N). For the \emph{W1} band, this relationship can be approximated as $3.0/(\rm{S/N}$) \citep{cut13:allwise,deb11:apjs197}. At the 5$\sigma$ detection limits of AllWISE, the astrometric uncertainty reaches 0.6\arcsec, meaning that in samples of a few hundred one reasonably expects several true detections of objects at separations greater than 0.5\arcsec. Conversely, and perhaps more detrimental, an object with high S/N within a separation of 0.5\arcsec\,could in fact be several standard deviations away from its expected position. Both cases emphasize that the raw separations should not be directly compared between bright and faint objects, and instead the individual astrometric uncertainty must be considered.

The framework developed for probability-based cross-matches provides a useful way to incorporate the astrometric uncertainty into the evaluation of whether or not the \emph{WISE} astrometric position is likely perturbed (see \cite{wil18:mnras481} for example). Additionally, the positional accuracy and proper motions provided by the \emph{Gaia} Data Release 2 \citep{gaia16:aap595,gaia18:aa616} provide a fantastic reference position. As part of the \emph{Gaia} DR2, cross-matched catalogs between several optical and near-infrared surveys were produced based on probabilistic, nearest-neighbor approaches \citep{mar17:aap607, mar19:aap621} that incorporate the astrometric uncertainty of each survey, the epoch differences between each catalog, and the probability of randomly finding a nearby, unrelated counterpart in a survey given the local source count density.

\begin{figure}
\epsscale{1.2}
\plotone{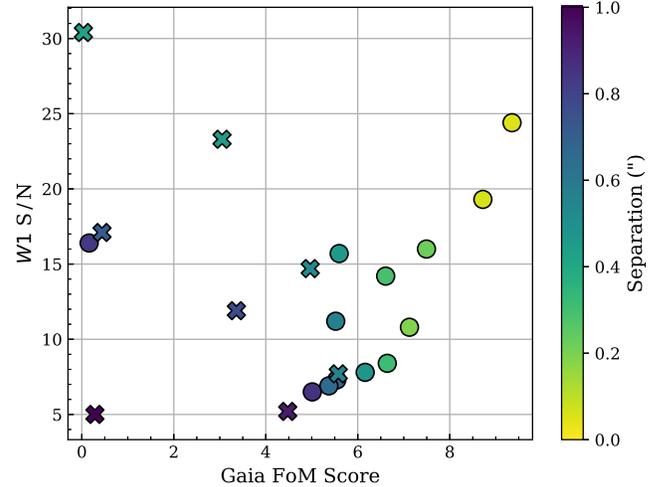}
\caption{The \emph{Gaia} Figure of Merit is plotted against the AllWISE S/N for each object, with confirmed excesses shown as circles and excesses due to \emph{WISE} source confusion as crosses. The color scale represents the separation between the expected position and the AllWISE detection. Candidates with a high \emph{W1} S/N but low FoM score are likely cases of confused AllWISE photometry. \label{fig:fomvssnr}}
\end{figure}

The cross-match algorithm works by first searching for all possible counterparts (dubbed neighbors) in a given catalog within 5$\sigma$ of the combined astrometric uncertainty of the object in \emph{Gaia} and the neighbors in the catalog of interest. The Figure of Merit (FoM) is computed for each potential neighbor by comparing the probabilities of discovery of the object at the measured separation and the probability of chance alignment. The counterpart with the highest FoM is selected as the match and reported in the \emph{bestNeighbor} table \citep{mar17:aap607}. All neighbors for each cross-match are listed in the corresponding \emph{Neighborhood} table. 

There is no threshold for the FoM score to use to evaluate the goodness of a match, that is to say the FoM does not translate directly into a likelihood. For the AllWISE catalog, this dimensionless parameter ranges from $7.0\times10^{-5}$ to 15.5 \citep{mar19:aap621}, with a strong dependence on the astrometric uncertainty of the counterpart in AllWISE. As the AllWISE astrometry is inversely proportional to the S/N ratio of the detection, one expects a relationship between the \emph{W1} S/N and the \emph{Gaia} FoM score. We queried the \emph{Gaia} \emph{Neighborhood} catalog for and collected the recorded separation and FoM score of the best neighbor identified in the \emph{Gaia} cross-match. 

Figure \ref{fig:fomvssnr} demonstrates a strong relationship between the \emph{Gaia} FoM score and the \emph{W1} S/N, where the majority of the outliers are cases of confused AllWISE photometry. Based on this, we conclude that excesses with S/N $>\,$10 but FoM $<\,$4 are likely the result of source confusion. There is one object in this region, ATLAS\,10552, that is a confirmed excess. A closer inspection of the images and SED for ATLAS\,10552 indicate it is the rare case where the AllWISE photometry was confused in addition to the white dwarf having a true infrared excess as there is a faint, nearby source and the IRAC fluxes are slightly below the AllWISE fluxes. 

Another object, EC\,21548, returned no neighbors in the Gaia-AllWISE cross-match, i.e. there is not an associated source to the \emph{Gaia} detection in the AllWISE catalog within 5$\sigma$ of astrometric separation. The AllWISE photometry we associated with EC\,21548 corresponds to a source found at a separation of 0.5\arcsec\,from the expected position of EC\,21548. The \emph{Spitzer} images show a single source near the expected position of EC\,21548, and the IRAC-Ch$\,$1 and Ch$\,$2 fluxes agree with the AllWISE \emph{W1} and \emph{W2} fluxes, leading to a bit of mystery as to why the \emph{Gaia} and the nearest AllWISE coordinates are so discrepant. Its exclusion in the cross-match could indicate some unaccounted for systematic uncertainty in the AllWISE astrometry, or it could simply be spurious. Whatever the case, it is another good example of a target that would have been erroneously rejected by the astrometric uncertainty cut proposed above. 

In addition to the two confirmed excesses that would have been rejected, a few cases of confused \emph{WISE} photometry are not distinguished by this method. SDSS\,00021, SDSS\,13054, and SDSS\,13570 all lie near the sample of confirmed infrared excesses. The first is a case of a statistically weak infrared excess, and can be discarded for the purpose of evaluating this technique. Referencing the \emph{Spitzer} images in Figure \ref{fig:imsequence}, we see that the remaining two have multiple sources contaminating the AllWISE photometry, resulting in a smaller positional perturbation than cases where one contaminant is responsible for the AllWISE positional offset. 

In general, the Gaia FoM is a useful discriminant for identifying confused \emph{WISE} photometry, having correctly identified five out of the eight confused sources in our sample. Applying this technique would have come at a cost though, as two confirmed excesses were rejected by this method and the two cases of multiple contaminants that result in little astrometric perturbation would have been missed. These results emphasize that even advanced astrometric methods will fail to produce clean samples of \emph{WISE}-selected infrared excesses.

\subsection{Proper Motion Comparison}

Related to the astrometric test, one can also compare the proper motions measured by \emph{WISE} and \emph{Gaia} to test the validity of a \emph{WISE}-selected infrared excess \citep{deb19:apjl872}. This is effectively repeating the astrometric experiment with a series of independent  measurements over time. Given the six month baseline, the initial \emph{WISE} proper motions are not sufficient for comparison with \emph{Gaia}, but the continued observations of the \emph{NEOWISE} mission \citep{mai14:apj792} have provided a six year baseline allowing for improved motion measurements. The CatWISE Preliminary catalog \citep{eis19:arxiv} provides new photomety and proper motion measurements using the original AllWISE processing techniques for data collected between 2010 and 2016, providing a factor of ten improvement to the original AllWISE proper motion measurements, in addition to improving the depth and positional accuracy of sources as compared to AllWISE.

The proper motion accuracy in CatWISE is 10 mas\,yr$^{-1}$ for bright sources, 30 mas\,yr$^{-1}$ at \emph{W1}\,$\approx$\,15.5 mag, and 100 mas\,yr$^{-1}$ at \emph{W1}\,$\approx$\,17 mag, so an object must either be sufficiently bright or have a sufficiently high proper motion to perform this test. Two of our objects meet this criterion, EC\,03103 and EC\,05276, and their reported proper motions are given in Table \ref{tab:pmtable}. Both objects have discrepant proper motions in \emph{Gaia} and CatWISE, consistent with their classification of having confused \emph{WISE} photometry. Unfortunately, the sample size is not sufficient to evaluate the efficacy of this technique, but the two cases of confirmed source confusion demonstrate that it is a worthwhile check for large surveys of \emph{WISE} infrared excesses. 

\begin{deluxetable}{lr}
\tablecaption{Comparison of \emph{Gaia} and CatWISE proper motion measurements for two candidates in our sample. All proper motion measurements are given in units of mas\,yr$^{-1}$\label{tab:pmtable}}
\tablehead{\multicolumn1l{EC\,03103} & \colhead{}}
\startdata
\emph{Gaia} Source ID & 4720876181720327808 \\
\emph{Gaia} DR2 $\mu_\alpha$\,cos($\delta$) & 404.8$\,\pm\,$0.2 \\
\emph{Gaia} DR2 $\mu_\delta$ & 57.6$\,\pm\,$0.2 \\
CatWISE Source Name & J031122.06-621515.2 \\
CatWISE $\mu_\alpha$\,cos($\delta$) & 279.9$\,\pm\,$21.2 \\
CatWISE $\mu_\delta$ & 80.3$\,\pm\,$19.3\\
\hline
\sidehead{EC\,05276}
\hline
\emph{Gaia} Source ID & 4805782462481529600 \\
\emph{Gaia} DR2 $\mu_\alpha$\,cos($\delta$) & -37.3$\,\pm\,$ 0.1 \\
\emph{Gaia} DR2 $\mu_\delta$ & 15.3$\,\pm\,$0.1 \\
CatWISE Source Name & J052912.09-430334.8 \\
CatWISE $\mu_\alpha$\,cos($\delta$) & -397.3$\,\pm\,$34.4 \\
CatWISE $\mu_\delta$ & 358.5$\,\pm\,$35.7\\
\enddata
\end{deluxetable}

\section{Conclusions}

Among the sample of 22 \emph{WISE}-selected dusty white dwarf candidates, we find that eight are the result of source confusion, despite our attempts at vetting the sample prior to \emph{Spitzer} observation. We show that ground-based, near-infrared imaging is insufficient for detecting the contaminants in our sample, but should still be employed when vetting candidates to rule out more obvious cases of source confusion. Astrometric filtering of candidates on the fainter end of the \emph{WISE} catalog should also take into account the astrometric uncertainty, and we demonstrate the utility of filtering candidates using the Figure of Merit metric from the official \emph{Gaia}-AllWISE cross-match. 

However, even when applying these techniques in combination one will fail to produce a clean sample of \emph{WISE}-selected infrared excesses, and care must be taken when interpreting the statistical properties of \emph{WISE}-selected infrared excesses. The fact remains that \emph{WISE}-selected infrared excess candidates should be treated as guilty until proven innocent. \deleted{and the best available facility capable of passing judgment is \emph{Spitzer}}\added{The confusion limit is inherent to the \emph{WISE} telescope and cannot be remedied by advanced processing. Future studies of \emph{WISE}-selected infrared excesses utilizing the new co-adds and increased depth of the continued NEOWISE mission \citep{sch19:apjs240} could suffer from even higher contamination rates, as the survey depth is pushed further and further past the confusion limit.}

The 14 confirmed excesses in our sample could also provide a nice increase to the known sample of dusty white dwarf stars, which currently stands between 40 and 50 systems \citep{far16:nar71}. We emphasize that our confirmation does not signify their status as dusty white dwarf stars, as we cannot preclude the possibility of a brown dwarf companion as the source of the infrared excess. To-date, all confirmed dusty white dwarf stars have also shown signs of active accretion detectable as atmospheric metals, and the search for these is a necessary step for solidifying their infrared excess as  circumstellar dust. Only one of the 14 \emph{Spitzer}-confirmed excesses in our sample has a literature detection of metals (EC\,01071; \citealt{den17:apj849}), and we are currently pursuing high resolution spectroscopic follow-up of the remaining candidates. 

\acknowledgments

We would like to acknowledge Boris G\"{a}nsicke for comments and suggestions which improved this manuscript, and the anonymous referee for providing a swift and helpful report. This work is based in part on observations made with the Spitzer Space Telescope, which is operated by the Jet Propulsion Laboratory, California Institute of Technology under a contract with NASA. This research has made use of the NASA/ IPAC Infrared Science Archive, which is operated by the Jet Propulsion Laboratory, California Institute of Technology, under contract with the National Aeronautics and Space Administration. This publication makes use of data products from the Wide-field Infrared Survey Explorer, which is a joint project of the University of California, Los Angeles, and the Jet Propulsion Laboratory/California Institute of Technology, and NEOWISE, which is a project of the Jet Propulsion Laboratory/California Institute of Technology. WISE and NEOWISE are funded by the National Aeronautics and Space Administration. This work has made use of data from the European Space Agency (ESA) mission {\it Gaia} (\url{https://www.cosmos.esa.int/gaia}), processed by the {\it Gaia} Data Processing and Analysis Consortium (DPAC, \url{https://www.cosmos.esa.int/web/gaia/dpac/consortium}). Funding for the DPAC has been provided by national institutions, in particular the institutions participating in the {\it Gaia} Multilateral Agreement.\\

\vspace{5mm}
\facilities{IRSA, Spitzer, WISE, Gaia}
\software{astropy \citep{ast13:aap558}}

\clearpage

\appendix
\restartappendixnumbering

\section{Spectral Energy Distributions\label{appendix}}

\begin{figure*}[ht!]
\epsscale{0.87}
\plotone{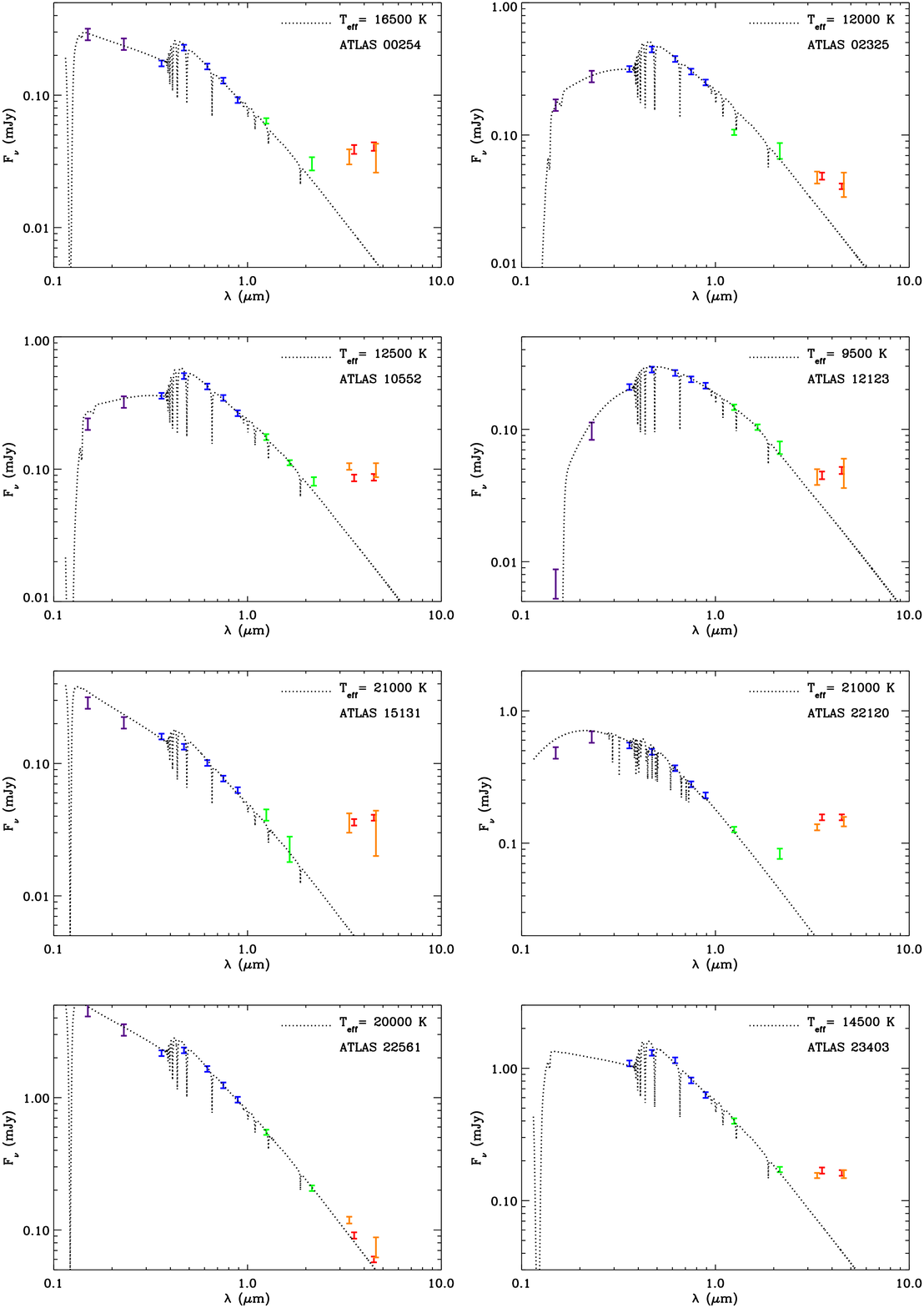}
\caption{Spectral energy distributions of remaining \emph{Spitzer} targets. The cases where the \emph{Spitzer} phtometry (red) is  consistent with the stellar models (dotten line) are targets with confused \emph{WISE} (orange) excesses.}
\end{figure*}

\begin{figure*}[ht!]
\figurenum{A1 continued}
\epsscale{0.87}
\plotone{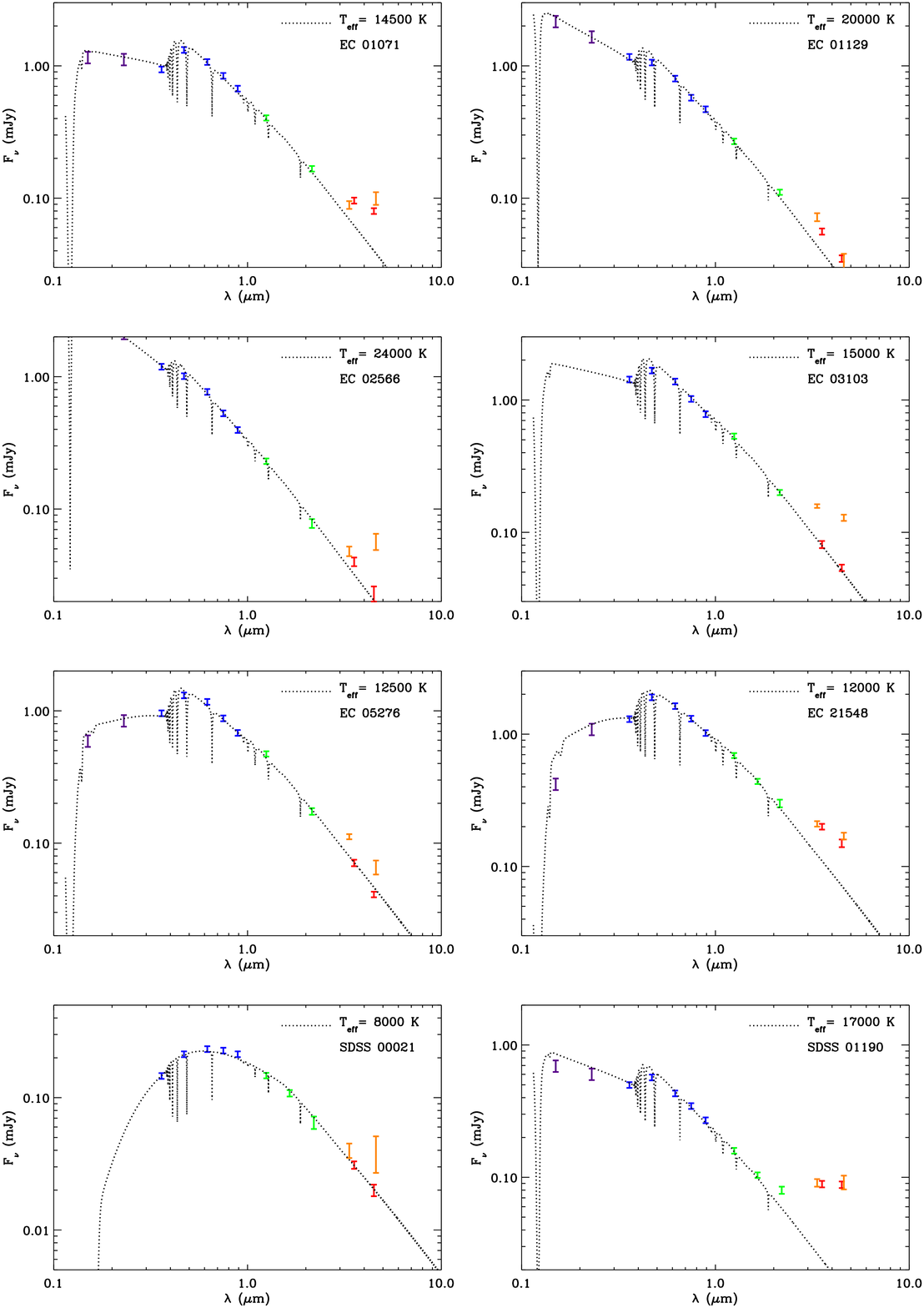}
\caption{}
\end{figure*}

\begin{figure*}[ht!]
\figurenum{A1 continued}
\epsscale{0.87}
\plotone{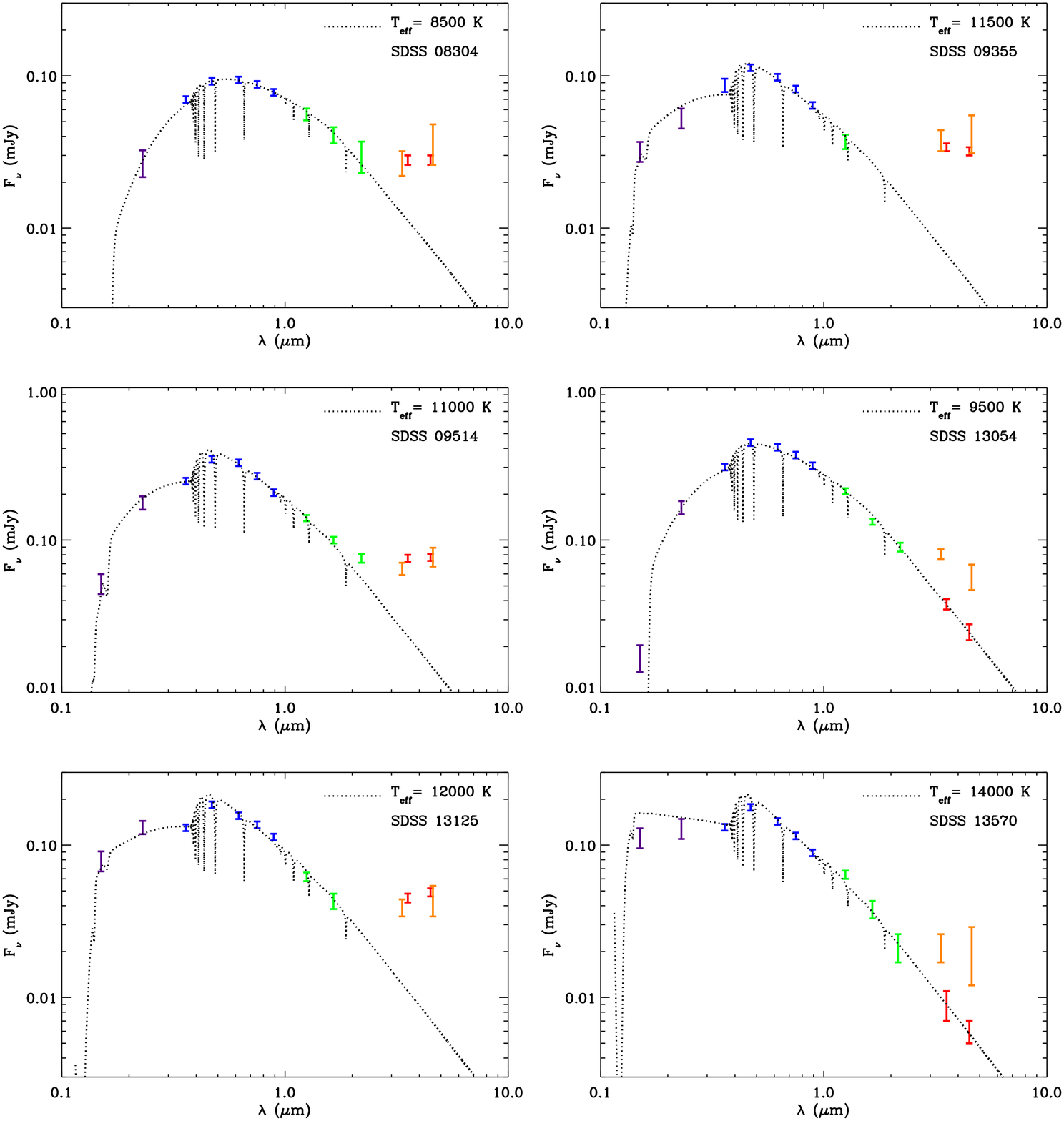}
\caption{}
\end{figure*}

\clearpage

\bibliography{wdexoplanets}
\bibliographystyle{aasjournal}

\end{document}